\documentclass[aps,pre,showkeys,twocolumn,showpacs,superscriptaddress]{revtex4-1}
\usepackage{amsmath,amssymb}
\usepackage{graphics,graphicx}
\usepackage{dcolumn,bm}
\usepackage{mathbbol}
\newcommand{\tcmp}
{\affiliation{Theoretical Condensed Matter Physics Division,
Saha Institute of Nuclear Physics, 1/AF Bidhannagar, Kolkata 700 064 India.}}

\begin{document}

\title{Mean field solutions of kinetic exchange opinion models}
\author{Soumyajyoti Biswas}%
\email[Email: ]{soumyajyoti.biswas@saha.ac.in}
\tcmp
%%%%%%%%%%%%%%%%%%%%%%%%%%%%%%%%%%%%%%%%%%%%%%%%%%%%%%%%%%%%%%%%%%%%%%%%%%%%%%%%%%%%%

\begin{abstract}
\noindent 
We present here the exact solution of an infinite range, discrete, opinion formation model. The model shows an active-absorbing phase transition,
similar to that numerically found in its recently proposed continuous version (Lallouache {\it et al.}, Phys. Rev E {\bf 82}, 056112 (2010)). 
Apart from the two-agent interactions here we also report the effect of having three agent interactions. The phase diagram
has a continuous transition line (two agent interaction dominated) and a discontinuous transition line (three agent interaction 
dominated) separated by a tricritical point. 
\end{abstract}

\maketitle

%%%%%%%%%%%%%%%%%%%%%%%%%%%%%%%%%%%%%%%%%%%%%%%%%%%%%%%%%%%%%%%%%%%%%%%%%%%%%%%%%%%%%
%%%%%%%%%%%%%%%%%%%%%%%%%%%%%%%%%%%%%%%%%%%%%%%%%%%%%%%%%%%%%%%%%%%
\section{Introduction}\label{sec:1}
%%%%%%%%%%%%%%%%%%%%%%%%%%%%%%%%%%%%%%%%%%%%%%%%%%%%%%%%%%%%%%%%%%%
\noindent The dynamics of opinion formation in a society and  emergence of consensus arising out of cooperative interactions between the agents are being 
extensively studied recently \cite{ESTP,Stauffer:2009,Castellano:RMP,Galam:1982,Liggett:1999,Sznajd:2000,Galam:2008, Sen:opin}. 
Due to its cooperative nature, statistical mechanics provides the required tools to study such systems.
 Although many intricacies of real societies are lost in the process, such minimal modeling often
yields essential features in terms of their  social as well as physical aspects.

The key feature in modeling opinion formation is to quantify opinions in terms of real numbers. Depending upon context,
 opinion is often quantified as discrete or continuous variables between two or more choices. 
Also the process of interaction between the agents is a vital ingredient. While several choices to model such interactions exist, 
one way is to consider an interaction as a `scattering process' where the agents are stochastically influenced by
each other's opinions (see, for example, \cite{Hegselman:2002,Deffuant:2000,Fortunato:2005,
Toscani:2006}).

Recently an opinion formation model \cite{Lallouache:2010} based on such a `kinetic exchange process' between two individuals was proposed [the Lallouache-Chakrabarti-Chakraborti-Chakrabarti (LCCC) model hereafter]. 
Resembling the model for wealth exchange in a society \cite{CC-CCM}, this model has a single parameter that determines the `conviction' of an individual. It was shown
that beyond a certain value of this conviction parameter, the society reaches a `consensus', where one of the two choices (positive or negative) provided 
to the individuals prevails, thereby spontaneously breaking a discrete symmetry. 

A generalization was proposed subsequently \cite{Sen:2010}, in which the `self-conviction' and the ability to influence others were taken as 
independent variables. This two-parameter model has a simple phase boundary along which apparent non-universality was observed
(for detailed discussion on the critical behavior of a class of model of this kind see \cite{bcc, eco}). 
Subsequent extensions in terms of including ``noise'' \cite{bcs} and also studying a generalized map version \cite{asc:2010}
for this class of models were done.

In the present paper, a discretised version of the LCCC model is exactly solved. As this is an infinite range model, the solution is essentially mean field.
Also a generalisation in  terms of three-agent interactions is proposed. From the expression of the
order parameter it is seen that for pure three-agent interactions (Model I) the transition is discontinuous (giving  hysteresis like behavior as well)
but for mixture of two-agent and three-agent interactions (Model II), the nature of transition depends on the relative probabilities of the two 
types (two-agents and three-agents) of interactions. 

As we will see, the present version of the model is a (three-state) probabilistic cellular automaton (PCA) (see, for example, \cite{dk1,dk2,dk3,Lubek,Hinrichsen:2000}).
Such types of models have been used in sociophysics for a long time starting from Schelling \cite{schelling} (see also \cite{nowak1, nowak2, flache, holyst} to name a few).

The paper is organized as follows: In Sec. II the existing (LCCC) and the proposed models will be described.
 In Sec. III the mean-field calculations to find the expression for
order parameter is presented. Then in Sec. IV the three-agent generalisation is introduced with the analysis of the order of transition.
In Sec. V, we discuss the phase boundary obtained for the model with both two-agent and three-agent interactions.
Finally, the results are discussed.  
%%%%%%%%%%%%%%%%%%%%%%%%%%%%%%%%%%%%%%%%%%%%%%%%%%%%%%%%%%%%%%%%%%%%
\section{Kinetic exchange opinion model and its discrete version}\label{sec:2}
%%%%%%%%%%%%%%%%%%%%%%%%%%%%%%%%%%%%%%%%%%%%%%%%%%%%%%%%%%%%%%%%%%%
\noindent Consider a simple pair-wise interaction model for opinion formation (LCCC) in a well-connected group of individuals.
The opinion of an individual in the society is represented by a real number which can continuously vary within the limit $-1\le o_i \le +1$.
At any time $t$ an agent with opinion $o_i(t)$ interacts with another randomly chosen agent with opinion $o_j(t)$. After the interaction
the $i$-th agent retains a fraction of her own opinion (which depends on the agent's `conviction') and is stochastically 
influenced by the $j$th agent. The amount of the influence, of course, depends upon the $j$th agent's `conviction'.
The dynamics of the LCCC model evolves following the equation
\begin{equation}
\label{lccc}
o_i(t+1)=\lambda_i o_i(t)+\lambda_j\epsilon o_j(t).
\end{equation}
The parameter $\lambda_m$ represents the conviction of $m$-th agent  and $\epsilon$ is a stochastic variable 
uniformly distributed between [0,1]. If the opinion of an agent reached the higher (lower) extreme $+1$ ($-1$), then of course its opinion 
value was prevented from further increase (decrease).  $N$ such exchanges (where $N$ is 
the total number of agents) constitute a single Monte Carlo (MC) time step. 
For simplicity, the society
was assumed to be homogeneous in the sense that all $\lambda_m$'s were same (say, $\lambda$). (Note that in Ref.\cite{Lallouache:2010}, simultaneous changes of opinions for
both the $i$th and $j$th agents were considered, whereas Eq. (\ref{lccc}) describes a single change.) 
As long as the dynamics is of infinite range (any agent can interact with any other), this difference does not affect the results, except for the fact that
the time is rescaled ($2N$ updates for a MC step).  

The steady-state characterization of this model was done using two measures. One is the average opinion of the agents
$O=\frac{1}{N}\sum\limits_{i=1}^No_i$ representing the measure of global consensus and the other is the fraction
of agents having extreme opinions 
\begin{equation}
C=f_1+f_{-1},  
\end{equation}
where $f_1$ and $f_{-1}$ are the fraction of agents having opinion value $+1$ and $-1$ respectively.

Extensive Monte Carlo (MC) study \cite{Lallouache:2010} yields that in the LCCC model, above $\lambda_c\approx\frac{2}{3}$, $O\ne 0$ and below $\lambda_c$,
$O=0$. Furthermore, in the disordered phase, which is rather special here, $o_i=0$ for all $i$. This can be considered as a neutral state of the agents
(avoiding such neutral states are possible in some cases \cite{bcs}).  Similar behavior was also obtained for $C$. As in usual critical phenomena, 
the relaxation time shows divergence
from both sides of the critical point following a power law, having same exponent value on either side of the criticality.
Although nothing could be predicted about the critical exponent values, a mean-field like analysis gave $\lambda_c=2/3$ for the LCCC model (for detailed discussions
see Ref.\cite{bcc}).

In its generalization \cite{Sen:2010}, it was argued that the `self-conviction' $\lambda$ of an agent need not, in general, be equal to her 
ability to `influence' others. In its generalized form, therefore, the dynamical exchange process reads
\begin{equation}
\label{ps}
o_i(t+1)=\lambda o_i(t)+\mu \epsilon o_j(t),
\end{equation}
where $\mu$ represents the $j$-th agents ability to influence others. In the limit $\lambda=\mu$ one recovers the LCCC model.
For this generalised model, there is a phase boundary in $\lambda-\mu$ plane, having the equation $\lambda_c=1-\mu_c/2$.
The values of the critical exponents along this phase boundary were reported to have strongly non-universal behavior \cite{Sen:2010} for $O$ and weakly non-universal
behavior for $C$.

The above-mentioned models defy  simple  analytical treatments to find the order parameter
as long as $o_i$'s are continuous. But it is often the case in a society that the opinions can take only discrete values (e.g., voting `yes' or `no'
for a referendum, or voting in a two-party political scenario etc.). 
While retaining the social interpretation, it significantly simplifies analytical treatment. To that end the
 following modifications are made in the present study:
 The dynamical 
exchange equation [Eq.(\ref{lccc})] remains the same, but we make $\lambda$ stochastic
in the sense that we put $\lambda=1$ with probability $p$ and $0$ with probability $1-p$. Also, the parameter $\epsilon$ is either $1$ or $0$ with equal probability.
This is a discretized version of the LCCC model where the agents have three possible opinion values (i.e., $o_i\in \{-1, 0, +1\} \forall i$). 
This is, therefore, a three-state probabilistic cellular automaton (PCA).

The above mentioned modifications lift the `homogeneous society' (all agents having same `conviction') assumption  
and on the other hand keep the opinion values discretised.
However, the inhomogeneity is the simplest of its kind: only `high'  and `low'
 convictions are present. The agents can change between these two states randomly in time ($\lambda$ is annealed variable). 
The case of quenched $\lambda$, in this case, is a trivial limit where order parameter becomes simply proportional to $p$.

In the case of the generalised version [Eq.(\ref{ps})] the additional change is that like $\lambda$, we put $\mu=1$ with probability
$q$ and $0$ with probability $1-q$.
However, regarding its variation in time, it is explicitly checked numerically throughout the paper that the results do not change 
 whether $\mu$ depends on time
or not [at least in the mean-field (MF) limit]. Therefore, to facilitate analytical treatment, it is assumed to be randomly varying with time. 

Finally note that the relevant parameters of the problem here will be $p$ and/or $q$, which essentially
specify the average values of $\lambda$ and $\mu$ respectively. Also, the present versions are three-state PCA, but one can prove (see Appendix A) that
in the steady state opinions of opposite signs do not coexist (both in `space' and time). Therefore, all the subsequent results will hold
for a two-state PCA as well.

%%%%%%%%%%%%%%%%%%%%%%%%%%%%%%%%%%%%%%%%%%%%%%%%%%%%%%%%%%%%%%%%%%%%%%
\section{Mean field solution of the discrete model}
\noindent It is our intention to find an expression for the order parameter $O$ in terms
of $p$ and to find out the order parameter exponent $\beta$ defined as $O\sim(p-p_c)^{\beta}$. Subsequently a similar expression in terms 
of both $p$ and $q$ for the generalised model (see Eq.(\ref{ps})) is also presented.

%%%%%%%%%%%%%%%%%%%%%%%%%%%%%%%%%%%%%%%%%%%%%%%%%%%%%%%%%%%%%%%%%%%%%%%%%%%%%%%%%%%%%%%%%%%%%%%%%%%%%%%%%%%%%%%%%%%%%%%%%%%%%%%%%%%%%%%%%
\begin{figure}[tb]
\centering \includegraphics[width=9cm]{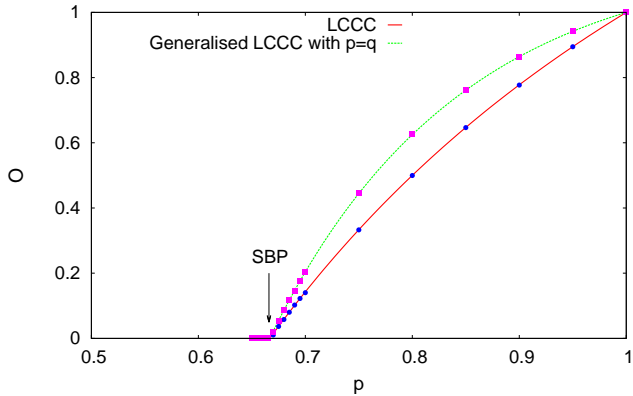}
   \caption{(Color online) Variation of the order parameter $O$ for discrete LCCC model and its generalised version (in the limit $p=q$). Analytical expressions given by Eqs. (\ref{op1})
(lower curve) and (\ref{op2}) (upper curve) are in good agreement with the simulation points. The critical or symmetry breaking point (SBP) $p_c=2/3$ is indicated. 
For the simulations, $N=10^5$ agents were considered, The initial conditions are ordered (all $+1$ or all $-1$). The results are time averaged (100 MC steps)
after the system reaches a steady states (roughly after $10^3$ initial MC steps) and about 100 ensemble averages (different starting configurations) were taken.}
\label{op1.2}
\end{figure}
%%%%%%%%%%%%%%%%%%%%%%%%%%%%%%%%%%%%%%%%%%%%%%%%%%%%%%%%%%%%%%%%%%%%%%%%%%%%%%%%%%%%%%%%%%%%%%%%%%%%%%%%%%%%%%%%%%%%%%%%%%%%%%%%%%%%%%%%%%

Let $f_0$, $f_1$ and $f_{-1}$ be the fractions of agents having opinions $0,+1$ and $-1$ respectively. Now, since the interactions
are only pair-wise and both $\lambda$ and $\mu$ can take only two values, one can enumerate all possible interactions between
all possible pairs, which contribute to increase and decrease of the order parameter. For example, the probability that
both the agents in an exchange process have opinion +1 is $f_1^2$. Now the probability with which one agent shifts her
opinion to 0 is $(1-p)$. Therefore the process $(1,1)\to(0,1)$ has probability $f_1^2(1-p)$. This process, of course, contributes
to decreasing of the order parameter. One can enumerate all the eight processes that contribute to changing the average order parameter ($O=f_1-f_{-1}$) and can 
therefore write the evolution  equation as

\begin{eqnarray}
&&\frac{dO}{dt}=f_{-1}^2(1-p)+f_{-1}f_1(1-\frac{p}{2})+\frac{f_0f_1p}{2}+f_{-1}f_0(1-p)\nonumber \\ &&-f_1^2(1-p)
-f_1f_{-1}(1-\frac{p}{2})-\frac{f_0f_{-1}p}{2}-f_{1}f_0(1-p)
\label{opdt}
\end{eqnarray}
In the steady state, the left hand side will be zero. Then after simplification one arrives at
\begin{eqnarray}
\label{ss1}
&& f_1^2(1-p)+f_0f_{-1}\frac{p}{2}+f_0f_1(1-p) \nonumber \\
&& =f_{-1}^2(1-p)+f_0f_1\frac{p}{2}+f_0f_{-1}(1-p)
\end{eqnarray}  
This gives either $f_1=f_{-1}$, (which implies disorder) or  
\begin{equation}
\label{f0}
f_0=\frac{2(1-p)}{p}.
\end{equation}

%%%%%%%%%%%%%%%%%%%%%%%%%%%%%%%%%%%%%%%%%%%%%%%%%%%%%%%%%%%%%%%%%%%%%%%%%%%%%%%%%%%%%%%%%%%%%%%%%%%%%%%%%%%%%%%%%%%%%%%%%%%%%%%%%%%%%%%%%
\begin{figure}[tb]
\centering \includegraphics[width=9cm]{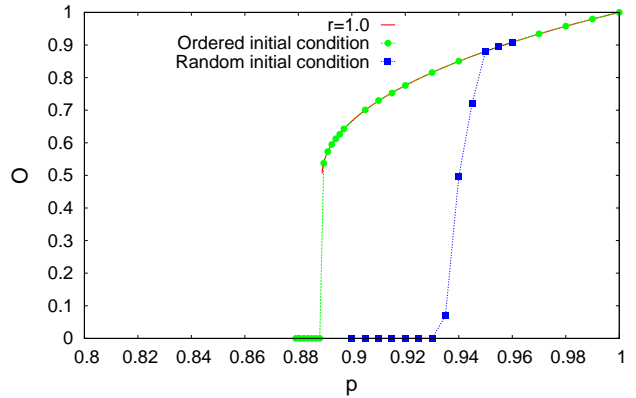}
   \caption{(Color online) Variation of the order parameter $O$ for pure three agent interactions. The solid line is given by Eq.~(\ref{op3}) i.e., the $r=1$
limit of Eq.~(\ref{opd}). In the simulation points, when the initial condition is an ordered state (all $+1$ or all $-1$),
the discontinuous jump occurs at $p_{c1}=8/9$, while starting from the random initial condition, the jump occurs at $p_{c2}\approx0.930\pm0.005$.
This clearly shows a hysteresis behavior (or, initial condition dependent behavior), which signifies the `resistance' of
the society to global changes in the opinion. $N=10^5$ agents are considered for the simulation points.}
\label{hys}
\end{figure}
%%%%%%%%%%%%%%%%%%%%%%%%%%%%%%%%%%%%%%%%%%%%%%%%%%%%%%%%%%%%%%%%%%%%%%%%%%%%%%%%%%%%%%%%%%%%%%%%%%%%%%%%%%%%%%%%%%%%%%%%%%%%%%%%%%%%%%%%%%

It is possible in this  case to show explicitly (see Appendix: A) that in the ordered state $f_1f_{-1}=0$ 
(making the steady state value of  $O$ and $C$ identical in this and for all subsequent discussions also). 
This condition and the disordered state condition ($f_1=f_{-1}$) should both be valid at the critical point. 
This is possible only when $f_1=f_{-1}=0$ at the critical point.  This implies, at the critical point $f_0=1$. 
Furthermore, for the sake of continuity of $f_1$
and $f_{-1}$, $f_0=1$ for the entire disordered phase.
This condition along with Eq.(~\ref{f0}) gives $p_c=2/3$. 

Therefore, the order parameter should be (using $f_1+f_{-1}+f_0=1$)
\begin{equation}
O=\pm(1-f_0)
\end{equation}
where the sign will depend on whether $f_1$ or $f_{-1}$ is non-zero in the ordered (symmetry-broken) phase. Using Eq.~(\ref{f0}), 
the above expression yields
\begin{equation}
\label{op1}
O=\pm\frac{3(p-\frac{2}{3})}{p}.
\end{equation}
Therefore, Eq.(\ref{op1}) gives $\beta=1$ (since $p_c=2/3$).

This can of course be generalized for Eq.(\ref{ps}). Straightforward algebra would yield 
\begin{equation}
f_0=\frac{(p-1)(q-2)}{pq}.
\end{equation}
As before, in the disordered phase $f_0=1$, which yields the equation for the phase boundary
in the $p$-$q$ plane as $p_c=1-\frac{q_c}{2}$. This gives the expression for order parameter as
\begin{equation}
\label{op2}
O=\pm\frac{2(p-p_c)+(q-q_c)}{pq}.
\end{equation}
Therefore, no matter through which path and which point the phase boundary is crossed, the order parameter exponent is $\beta=1$. 
The discretized version of the LCCC model 
presented here  has a transition, which is mean-field active-absorbing type \cite{Lubek,Hinrichsen:2000}. Accordingly $\beta=1$ is obtained. 

Of course we do not expect to get Eq.(\ref{op1}) from Eq.(\ref{op2}) by putting $p=q$, as this would only mean $\langle\lambda\rangle=\langle\mu\rangle$
and not $\lambda=\mu$.

The above-mentioned results are verified using numerical simulations. Ordered initial conditions (all $+1$ or all $-1$) were taken. The system was
allowed to relax until steady state was reached (checked by monitoring the time variation of the order parameter). For $N=10^5$, $10^3$ initial time
steps were discarded for relaxation (depending on the proximity of the model parameters $p,q$ with their critical values, this has to be increased upto $10^4$
for some of the points presented). 
Then a time average was taken for about 100 MC steps (1 MC step being N update attempts). The whole process was averaged over 100 ensembles.
Note that a random initial condition ($+1,-1,0$ with equal density) will give same steady-state values for the order parameter. But the relaxation time
will be higher.  

In Fig.~\ref{op1.2},  Eq.~(\ref{op1}) is compared with Monte Carlo simulations to find good agreement.
%%%%%%%%%%%%%%%%%%%%%%%%%%%%%%%%%%%%%%%%%%%%%%%%%%%%%%%%%%%%%%%%%%%
%%%%%%%%%%%%%%%%%%%%%%%%%%%%%%%%%%%%%%%%%%%%%%%%%%%%%%%%%%%%%%%%%%
\section{Model I: Beyond pair-wise interactions: three-agent interaction and discontinuous transition}
\noindent In all previous studies  \cite{Lallouache:2010, Sen:2010, bcc} regarding the kinetic exchange processes mentioned here, 
interactions were always taken to be occurring between 
two agents. This is partly because two-body exchange is the simplest and also because in the energy exchange of ideal gas 
only two-body interactions are important. But in opinion formation, exchange between more than two agents is perfectly possible.
So we intend to investigate the effect of such interactions in opinion formation.

The simplest possible generalisation toward many-body interaction is to consider three-body exchange. In doing so, the
following strategy is followed. Three agents are chosen randomly. Then one agent modifies her opinion according to that of
the other two only when the other two agrees among themselves. If they do not agree the first agent considers the group to be `neutral'.
Mathematically this can be represented as
\begin{equation}
\label{3body}
o_i(t+1)=\lambda o_i(t) + \lambda\epsilon\theta_{jk}(t), 
\end{equation} 
where,
\begin{equation}
\theta_{jk}(t)= \begin{cases} o_j(t) &\text{if  $o_j(t)=o_k(t)$}  \\
 0 &\text {otherwise.} 
\end{cases}
\end{equation}

%%%%%%%%%%%%%%%%%%%%%%%%%%%%%%%%%%%%%%%%%%%%%%%%%%%%%%%%%%%%%%%%%%%%%%%%%%%%%%%%%%%%%%%%%%%%%%%%%%%%%%%%%%%%%%%%%%%%%%%%%
\begin{figure}[tb]
\centering \includegraphics[width=9cm]{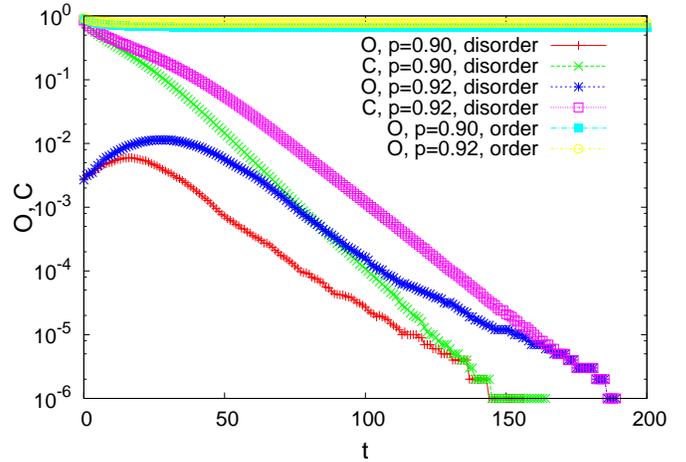}
   \caption{(Color online) The time dependence of $O$ and $C$ are plotted for two different values of $p$ ($0.90, 0.92$) both in the range $p_{c1}>p>p_{c2}$ 
for random ($o_i=1$ and $-1$ with equal probability) and fully ordered ($o_1=1$ for all $i$) initial conditions in the pure three agent interaction case.
 $O$ and $C$ are identical by definition 
for the ordered initial condition. For the random initial condition, it is seen that both $C$ and $O$ go to zero exponentially (linear in log-linear scale) confirming
the initial condition dependent behavior in this region of $p$ values. Number of agents is $N=10^5$ in the simulations.}
\label{time}
\end{figure}
%%%%%%%%%%%%%%%%%%%%%%%%%%%%%%%%%%%%%%%%%%%%%%%%%%%%%%%%%%%%%%%%%%%%%%%%%%%%%%%%%%%%%%%%%%%%%%%%%%%%%%%%%%%%%%%%%%%%%%%%%

Before proceeding further, it is to be noted that $\theta_{jk}(t)$ takes the value $+1,-1$ and $0$ with probabilities $f_1^2$, $f_{-1}^2$
and $1-(f_1^2+f_{-1}^2)$ respectively. Then just as Eq.(\ref{lccc}) was treated, one can enumerate all exchange processes that contribute to
increase and decrease in the order parameter. The time evolution equation is
\begin{eqnarray}
&&\frac{dO}{dt}=f_{-1}^3(1-p)+f_{-1}f_1^2(1-\frac{p}{2})+\frac{f_0f_1^2p}{2}\nonumber \\
&&+f_{-1}\left[1-(f_1^2+f_{-1}^2)\right](1-p)-f_1^3(1-p)-f_1f_{-1}^2(1-\frac{p}{2})\nonumber \\
&&-\frac{f_0f_{-1}p}{2}-f_1\left[1-(f_1^2+f_{-1}^2)\right](1-p)
\end{eqnarray}

Again, in the steady state left hand side should be zero, giving 
\begin{eqnarray}
&&f_1^3(1-p)+f_1f_{-1}^2\left(1-\frac{p}{2}\right)+f_0f_{-1}^2\frac{p}{2}\nonumber \\
&&+f_1\left[1-(f_1^2+f_{-1}^2)\right](1-p) \nonumber \\
&&=f_{-1}^3(1-p)+f_{-1}f_1^2\left(1-\frac{p}{2}\right)+f_0f_1^2\frac{p}{2}+\nonumber \\
&&f_{-1}\left[1-(f_1^2+f_{-1}^2)\right](1-p).
\end{eqnarray}
This gives either $f_1=f_{-1}$ (which implies disorder) or in the ordered state 
%%%%%%%%%%%%%%%%%%%%%%%%%%%%%%%%%%%%%%%%%%%%%%%%%%%%%%%%%%%%%%%%%%%%%%%%%%%%%%%%%%%%%%%%%%%%%%%%%%%%%%%%%%%%%%%%%%%%%%%%%
\begin{figure}[tb]
\centering \includegraphics[width=9cm]{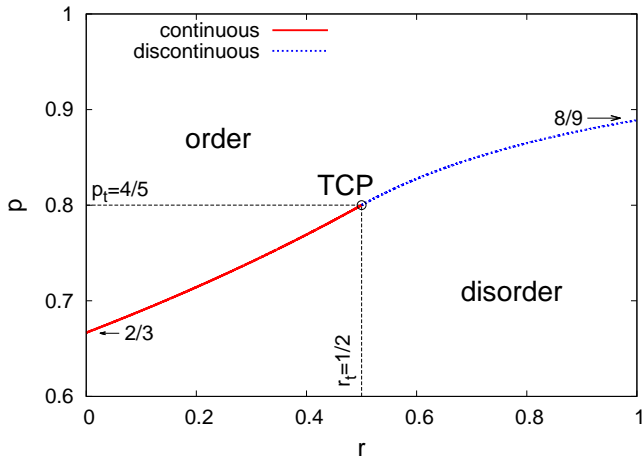}
   \caption{(Color online) The phase diagram of the model with mixture of three agent and two agent interactions in the $p$-$r$ plane, where
$r$ denotes the fraction of three agent interactions. The continuous transition  line follows Eq.(\ref{pb1}), while the discontinuous transition line follows Eq.(\ref{pb2}). Clearly, in the limits $r=0$ and $1$ the transition points are $2/3$ and $8/9$ respectively, as is expected from the discussions
in the text.}
\label{phdia}
\end{figure}
%%%%%%%%%%%%%%%%%%%%%%%%%%%%%%%%%%%%%%%%%%%%%%%%%%%%%%%%%%%%%%%%%%%%%%%%%%%%%%%%%%%%%%%%%%%%%%%%%%%%%%%%%%%%%%%%%%%%%%%%%
\begin{equation}
f_0=\frac{1}{2}-\frac{3\sqrt{p-8/9}}{2\sqrt{p}},
\end{equation}
where we have used $f_1f_{-1}=0$ as a simplifying assumption which is numerically verified here. 
We have neglected one solution of $f_0$ in which it increases with respect to $p$ in the ordered phase (because in the limit $p=1$, we expect $f_0=0$).
Using this, the order parameter takes the form
\begin{equation}
\label{op3}
O=\pm\left(\frac{1}{2}+\frac{3\sqrt{p-8/9}}{2\sqrt{p}}\right).
\end{equation}

Clearly, the above equation gives real values for $O$ only when $p>8/9$. Therefore, for $p<8/9$ the only real solution can be $f_1=f_{-1}$ i.e., $O=0$.
But from the form of Eq.(\ref{op3}) it is clear that in the ordered phase, the minimum value of $O$ can be 1/2. Therefore, the order-disorder transition
is necessarily discontinuous.

To verify this and to see the initial condition dependent behavior, MC simulations were performed. Depending
on the initial condition, the discontinuous jump from order to disorder happen at two different points, thus showing hysteresis behavior (see Fig.\ref{hys}). 
When the 
initial condition is ordered, Eq.(\ref{op3}) is followed upto $p_{c1}=8/9$. After that the order parameter jumps to zero. On the other hand, when the initial 
condition is random (having almost equal number of agents having opinions of opposite signs) then $O=0$ upto $p_{c2}\approx 0.930\pm0.005$ and then suddenly
jumps to the ordered (symmetry broken) phase. Of course $p_{c2}$ is the symmetry breaking point. Note that the estimation of $p_{c2}$ is entirely numerical here.
Fig.~{\ref{time}} shows the time dependence of $O$ and $C$ in the region where $p_{c1}>p>p_{c2}$. It is seen that for random initial condition, both $C$ and $O$
go to zero. This signifies that $o_i=0$ for all $i$ in this case, while those curves for the ordered initial condition shows saturation to a non-zero value.
This clearly confirms the initial condition dependent behavior in this region of $p$ values. 
For one particular point ($p=0.91$) this is checked for large enough size ($N=10^6$) for two different initial
conditions. It is clearly seen then the long time saturation values are very different, as is expected.

This initial-condition-dependent behavior  can be somewhat understood as follows: 
When the initial condition is random, the two agents chosen for the interaction with the first one have 
rather smaller chance of agreement among themselves. So the first agent sees the group as neutral (i.e., with opinion value zero).
In this way, the agent finds itself in an environment that effectively has a lot of neutral agents. Therefore, unless the agent's
conviction is very high (turns out to be about $0.93$), all of them succumb to this neutral environment and an absorbing phase is reached.
The situation, however, is very different when the initial condition is ordered. Here the environment is rather supportive in maintaining 
the prevailing opinion in the society (as is generally seen). Therefore, intuitively we expect that only at a lower  
value of conviction, the prevailing opinion will decay. This gives rise to the initial-condition-dependent behavior. 
%%%%%%%%%%%%%%%%%%%%%%%%%%%%%%%%%%%%%%%%%%%%%%%%%%%%%%%%%%%%%%%%%%%%%%%%%%%%%%%%%%%%%%%%%%%%%%%%%%%%%%%%%%%%%%%%%%%%%%%%%
\begin{figure}[tb]
\centering \includegraphics[width=9cm]{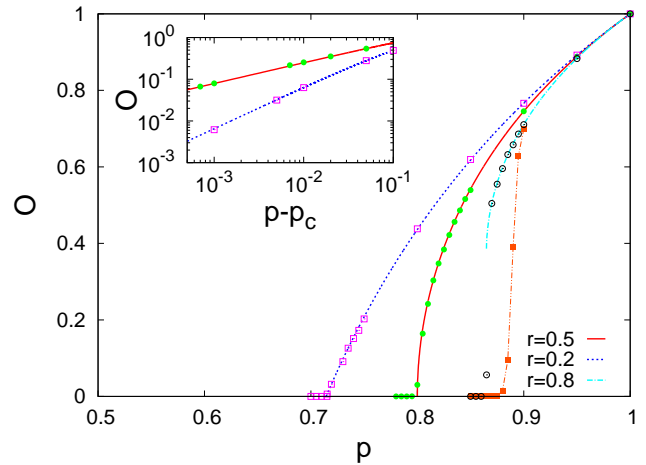}
   \caption{(Color online) Variations of the order parameter for  $r=0.2 (<r_t)$,  $r=0.5 (=r_t)$ and  $r=0.8 (>r_t)$. The continuous lines are analytical results (see Eq.(\ref{opd})) and the points are simulation results with $N=10^5$. For the first two curves, continuous
transitions are seen, as expected. For the last one ($r=0.8$) a discontinuous transition with signature of hysteresis is seen. Inset shows
the log-log plots of $O$ versus $p-p_c$ near the transition points for the continuous transitions. From the slopes of the curves the order parameter exponent $\beta$ is found to be $1$ and $1/2$ for $r=0.2$ and $r=0.5$ respectively.}
\label{many}
\end{figure}
%%%%%%%%%%%%%%%%%%%%%%%%%%%%%%%%%%%%%%%%%%%%%%%%%%%%%%%%%%%%%%%%%%%%%%%%%%%%%%%%%%%%%%%%%%%%%%%%%%%%%%%%%%%%%%%%%%%%%%%%%

Initial-condition-dependent behavior in opinion formation has been studied before in different contexts. 
For example, in Ref. \cite{holyst}, `hysteresis' was observed while modeling
the influence of a strong leader in the society. In general, these behaviors signify the resistance offered by the society to  changes in the
global opinion (both from order as well as disordered state), despite the fact that the very reason for its formation has  lost its relevance. 
In the present case too, the `hysteresis' (this word here does not have the usual meaning, say, that in the magnetic system) 
loop area is somewhat a measure of this social tolerance and resistance. 

A similar approach can be made for the generalised case described by Eq.(\ref{ps}). However, instead of doing so, one could also look at another
limit of Eq.(\ref{ps}), where $q=1$ (studied in Ref. \cite{bcc} as Model C). One can show that even in this limit, a discontinuous transition can be
obtained with $p_{c1}=(2+\sqrt{2})/4$. So one would in general expect a discontinuous transition for all ranges of Eq.(\ref{ps}).
%%%%%%%%%%%%%%%%%%%%%%%%%%%%%%%%%%%%%%%%%%%%%%%%%%%%%%%%%%%%%%%%%%
\section{Model II: Mixture of two-agent and three-agent interactions: Phase diagram and tri-critical point}
\noindent Let us now discuss how robust is this discontinuous transition. In the above analysis it was assumed that only three-agent interactions are 
present as opposed to the previous cases, where only two-agent interactions were considered. Here we consider a situation where both two-agent and
three-agent interactions are allowed. In principle interactions of all sizes should be allowed, but this is the simplest generalisation one could make.

With probability $r$ an exchange process is three-agent and otherwise it is two-agent. The exchange equation is the same as Eq.(\ref{3body}) but now clearly
\begin{equation}
\theta_{jk}(t)= \begin{cases} 1 & \text{with probability  $rf_1^2+(1-r)f_1$}  \\
 -1 & \text{with probability  $rf_{-1}^2+(1-r)f_{-1}$}  \\
 0 & \text{otherwise.}
\end{cases}
\end{equation}  
With this, one may enumerate all possibilities of increase and decrease of the order parameter and the time evolution equation will be
\begin{eqnarray}
&&\frac{dO}{dt}=f_{-1}\left[rf_{-1}^2+(1-r)f_{-1}\right](1-p)\nonumber \\
&&+f_1\left[rf_{-1}^2+(1-r)f_{-1}\right](1-\frac{p}{2})+\frac{f_0p}{2}\left[rf_1^2+(1-r)f_1\right] \nonumber \\
&& +  f_{-1}\left[1-(f_1^2+f_{-1}^2)r-(1-r)(f_1+f_{-1})\right](1-p) \nonumber \\
&& - f_1\left[rf_1^2+(1-r)f_1\right](1-p) \nonumber \\
&& -f_{-1}\left[rf_1^2+(1-r)f_1\right](1-\frac{p}{2})-\frac{f_0p}{2}\left[rf_{-1}^2+(1-r)f_{-1}\right]\nonumber \\
&& -f_1\left[1-(f_1^2+f_{-1}^2)r-(1-r)(f_1+f_{-1})\right](1-p).
\end{eqnarray}
In the steady state, the left hand side will be zero, giving (assuming $f_1f_{-1}=0$, which can be verified numerically) 
\begin{eqnarray}
&& f_1\left[rf_1^2+(1-r)f_1\right](1-p)+\frac{f_0p}{2}\left[rf_{-1}^2+(1-r)f_{-1}\right] \nonumber \\
&& +f_1\left[1-(f_1^2+f_{-1}^2)r-(1-r)(f_1+f_{-1})\right](1-p) \nonumber \\
&& = f_{-1}\left[rf_{-1}^2+(1-r)f_{-1}\right](1-p)+\frac{f_0p}{2}\left[rf_1^2+(1-r)f_1\right] \nonumber \\
&& +f_{-1}\left[1-(f_1^2+f_{-1}^2)r-(1-r)(f_1+f_{-1})\right](1-p).
\end{eqnarray}
On simplification, this yields either $f_1=f_{-1}$ which implies disorder, or
\begin{eqnarray}
\frac{pr}{2}f_0^2-\frac{pf_0}{2}+1-p=0,
\end{eqnarray}
which gives (the only relevant solution)
\begin{eqnarray}
f_0=\frac{1}{2r}-\frac{\sqrt{p^2/4-2pr(1-p)}}{pr}.
\end{eqnarray}
Again as before
\begin{eqnarray}
\label{opd}
O=\pm\left(\frac{2r-1}{2r}+\frac{\sqrt{p^2/4-2pr(1-p)}}{pr}\right).
\end{eqnarray}
The sign is determined from the fact whether $f_1$ or $f_{-1}$ is non-zero in the ordered phase.
However, the terms inside the bracket essentially represent a probability ($1-f_0$) and therefore, should be a positive and real number.
Now, when $r<1/2$ for the quantity to become positive, $p$ has to be sufficiently high ($>p_c$, say). Before that the disordered solution
($O=0$) remains valid. Therefore, the ordered-disordered phase is separated by a line in the $p-r$ plane, which is obtained by setting
the quantity inside the bracket to zero. That critical line is
\begin{equation}
\label{pb1}
 p_c=\frac{2}{3-r_c} \qquad \mbox{for} \quad r<\frac{1}{2},
\end{equation} 
where $p_c,r_c$ are the set of values satisfying the above equation where a continuous transition takes place.
Note that the above condition also keep $O$ real, which is the other requirement, as long as $r<1/2$.
In Eq.(\ref{opd}) one can put $p=2/(3-r)+\Delta$ (where $\Delta\to 0$) and show that the leading order term
comes out to be linear in $\Delta$, implying $\beta=1$ along this line. 
This critical line, of course, terminates at $(p_t=\frac{4}{5},r_t=\frac{1}{2})$. 

When $r>1/2$, the ordered-state solution Eq. (\ref{opd}) can be valid whenever it gives real
values for $O$. For this, the phase boundary is obtained by setting the quantity inside the square-root to zero. The said phase boundary is
\begin{equation}
\label{pb2}
p_c=\frac{8r_c}{1+8r_c} \qquad \mbox{for} \quad r>\frac{1}{2}.
\end{equation}

When $r>1/2$, the minimum value possible for $O$ from the ordered state solution (Eq.(\ref{opd})) 
is greater then zero. Therefore transition across this line is necessarily discontinuous.
This discontinuous nature is verified numerically. The amount
of discontinuity, of course, is given by $1-\frac{1}{2r}$, which is maximum (1/2) for pure three agent interactions ($r=1$). 
 A `hysteresis' like behavior, as discussed in the preceding section, is also seen.
Note that the phase boundary equations correctly 
give $p_c=2/3$ and $p_c=8/9$ limits respectively for $r=0$ (from Eq.(\ref{pb1})) and $r=1$ (from Eq.(\ref{pb2})).

The point $(p_t=\frac{4}{5},r_t=\frac{1}{2})$ is special where the critical line terminates. 
It is a Tricritical Point (TCP). As is seen generally, at the TCP the exponent values are different. Clearly, 
\begin{equation}
O\sim \sqrt{p-4/5}
\end{equation}  
giving $\beta_{TCP}=1/2$, which is different from $\beta=1$ found along the critical line. 

The phase boundaries and the tri-critical point are shown in Fig.~\ref{phdia}. In Fig.~\ref{many}, the order parameters are plotted
for three different values of $r$, one on either side of the TCP and one at TCP. The inset shows the different values for the
exponent $\beta$ when the $r<r_t$ ($\beta=1$) and when $r=r_t$ ($\beta=1/2$).

\begin{figure}[tb]
\centering \includegraphics[width=9cm]{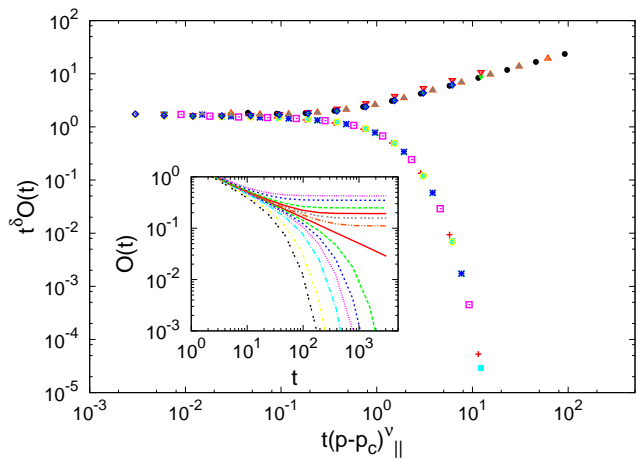}
   \caption{(Color online) Data collapse for finding $\nu_{\parallel}$ (see Eq.(\ref{scale})) for different $p$ values for $r=0.5$ ($p_c=0.5$). 
The estimate is $\nu_{\parallel}=1.00\pm0.01$. Inset shows the uncollapsed data. $N=10^5$ agents are considered, initial condition is ordered (all $+1$ or all $-1$).}
\label{nu}
\end{figure}

To find the other exponent values that characterize this TCP, one can use the following scaling relation for the order parameter
\begin{equation}
\label{scale}
O(t)\approx t^{-\delta}\mathcal{F}\left(t^{1/\nu_{\parallel}}\Delta, t^{d/z}/N \right),
\end{equation} 
where $\Delta=p-p_c$, $\nu_{\parallel}$ is the time-correlation exponent, $z$ is the dynamical exponent and $d$ is the space dimension. At the critical point, the order parameter
follows a power-law relaxation $O(t)\sim t ^{-\delta}$ (see inset of Fig.~\ref{nu} ) with $\delta=0.50\pm 0.01$. 
%%%%%%%%%%%%%%%%%%%%%%%%%%%%%%%%%%%%%%%%%%%%%%%%%%%%%%%%%%%%%%%%%%%%%%%%%%%%%%%%%%%%%%%%%%%%%%%%%%%%%%%%%%%%%%%%%%%%%%%%%%%%%%%%%%%%%%%%%
%%%%%%%%%%%%%%%%%%%%%%%%%%%%%%%%%%%%%%%%%%%%%%%%%%%%%%%%%%%%%%%%%%%%%%%%%%%%%%%%%%%%%%%%%%%%%%%%%%%%%%%%%%%%%%%%%%%%%%%%%%%%%%%%%%%%%%%%%%

%%%%%%%%%%%%%%%%%%%%%%%%%%%%%%%%%%%%%%%%%%%%%%%%%%%%%%%%%%%%%%%%%%%%%%%%%%%%%%%%%%%%%%%%%%%%%%%%%%%%%%%%%%%%%%%%%%%%%%%%%%%%%%%%%%%%%%%%%
\begin{figure}[tb]
\centering \includegraphics[width=9cm]{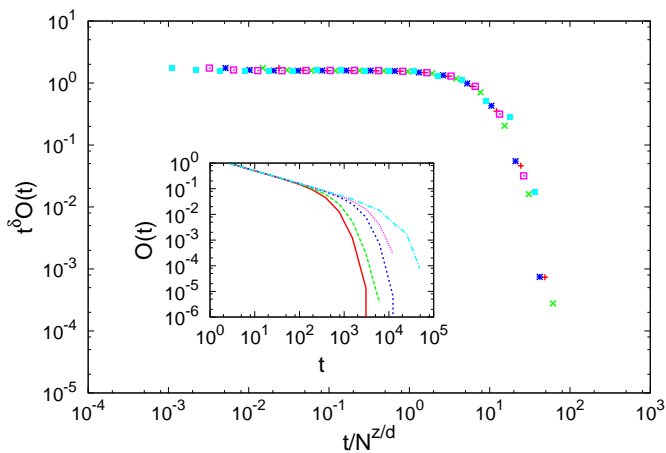}
   \caption{(Color online) Data collapse for finding $z$ for different system sizes ($N=500,1000,5000,10000,50000$) at $p=0.5$ and $r=0.5$ (TCP). 
The estimate is $z/d=0.666 \pm 0.001$. $N=10^5$ agents are considered, initial condition is ordered (all $+1$ or all $-1$). Inset shows the uncollapsed data.}
\label{tcpz}
\end{figure}
%%%%%%%%%%%%%%%%%%%%%%%%%%%%%%%%%%%%%%%%%%%%%%%%%%%%%%%%%%%%%%%%%%%%%%%%%%%%%%%%%%%%%%%%%%%%%%%%%%%%%%%%%%%%%%%%%%%%%%%%%%%%%%%%%%%%%%%%%%
One could then plot $O(t)t^{\delta}$ against $t(p-p_c)^{\nu_{\parallel}}$. By knowing $\delta$,
$\nu_{\parallel}$ can be tuned to find data collapse. From Fig.~\ref{nu} the estimate of $\nu_{\parallel}$ is $1.00\pm 0.01$. Similarly, 
one can plot $O(t)t^{\delta}$ against $t/N^{z/d}$. Again by
tuning $z$, data collapse is found (see Fig.~\ref{tcpz}). The estimate of $z/d$ is $0.666\pm 0.001$. 
To find $z$ one should put $d=4$, which is the upper critical dimension. This gives $z\approx 8/3$. Similar analysis for $r<\frac{1}{2}$ gives $z/d\approx1/2$,
here also by putting $d=4$ one gets $z\approx 2$, which is expected for directed percolation (DP).

The scaling relation $\delta=\beta/\nu_{||}$ is clearly satisfied here. 

Therefore we see that the nature of the transition is actually determined by the relative probabilities of the two-agent and three-agent interactions. Also the
exponent values at the tri-critical point are different from those along the critical line.
%%%%%%%%%%%%%%%%%%%%%%%%%%%%%%%%%%%%%%%%%%%%%%%%%%%%%%%%%%%%%%%%%%%%555
\section{Discussion}
\noindent In the first part of this paper an exact solution of the discretised version of a recently proposed model (LCCC) \cite{Lallouache:2010} for opinion dynamics is given.
The model being infinite ranged, the solution is essentially mean field.
The interactions, as in its simplest form, are between two agents [see Eq.(\ref{lccc})]. The exchange process is such that an agent has a `conviction' with which
she retains her opinion and also gets influenced (stochastically, because it is otherwise impossible to incorporate all social complexities involved in such
processes) by the opinion of one randomly chosen agent. It was shown \cite{Lallouache:2010, Sen:2010} from extensive MC study that beyond a certain value of the `conviction' 
parameter the society undergoes a phase transition from disordered to ordered state (where consensus is formed).
 In the present study that behavior is shown analytically [see Eq.(\ref{op1})] for a discretized version of the model in mean-field limit (which is exact here).
The order-parameter exponent has been found to be $\beta=1$. Even for the generalised version \cite{Sen:2010} [see Eq.(\ref{ps})] 
this exponent remains the same along the phase boundary (belonging to DP universality class).

Thereafter, a generalisation of this model for three-agent interaction is reported (Model I). There is, of course, no single choice for this kind of generalisation. 
But here we have
taken a plausible strategy in which an agent can be influenced by the opinions of two other randomly chosen agents only when those two agents agree
 among themselves (have the same opinion)
otherwise the first agent considers the group as `neutral' [see Eq.(\ref{3body})]. 
This generalisation
has led to an interesting behavior in terms of the order of the transition. It is seen if all interactions are three-agent, a discontinuous transition is obtained 
[see Eq.(\ref{op3})]
and an initial condition dependent behavior was also observed (Fig.\ref{hys}). It is to be noted that `hysteresis'-like behavior 
in opinion models have been reported before in other 
contexts (see e.g., Refs. \cite{holyst,herrmann}). 
In general, this behavior in opinion dynamics models
can be taken as a signature of the tolerance of the society, or in other words, its resistance to  changes in global opinion (as is also indicated in Ref. \cite{holyst}).  
Although a direct correspondence to a measurable quantity cannot be made from these simplified models, qualitatively this
 loop-area is somewhat a `measure' of this social `tolerance' mentioned above.

It is important to find out how much of this discontinuous nature is generic or it is  an artifact of the restriction of only three-agent interaction, as invoked by
Eq. (\ref{3body}).
Of course it is not possible (or at least it is very difficult) to allow interactions of all sizes as it should be in a real society. But at the very least one can
allow both two-agent and three-agent interactions with some probabilities (Model II). In doing so it is found that up to the point when the probability of three-agent
interactions is below $1/2$, the transition is continuous [phase boundary given by Eq.(\ref{pb1})] and beyond that the transition is discontinuous,
[phase boundary is given by Eq.(\ref{pb2}]. The point where the two-agent and three-agent interactions are equally probable, is a special point, because
it is a tri-critical point. The transition here is continuous, but the values of the exponents are different from those along the critical line. Along the critical line,
the exponents are of course mean-field active-absorbing type ($\beta=1$, $z=2$, $\nu_{||}=1$, $\delta=1$) but at the TCP they are 
$\beta = 1/2$, $z/d\approx .666 \pm 0.001 \approx 2/3$ 
($z\approx 8/3$), 
$\delta=0.50\pm0.01\approx 1/2$, 
$\nu_{||}\approx 1.00 \pm 0.01$. 

At this point it is appropriate to mention that the `disordered' phase in all these versions is quite special in the sense that all the agents are neutral in this phase.
One can avoid such situation (see Ref. \cite{bcs}) and make the `disordered phase' have coexistence of opinions of different signs. But here such generalisations
were not discussed. Even without such complexities, which one can add anyway with this, one finds intriguing features in this model.

One may note that while attempting to interpolate between a continuous and discontinuous transition, a tri-critical point was obtained also in
Ref. \cite{ep-herrmann}. There too, the tricritical point was situated at the point where the phase boundary changed its curvature.

Finally, one must also note that in all the above cases, the  existence  of a phase transition has not been proved. The two solutions representing 
ordered and disordered states were analytically obtained. The existence of the phase transition, for all practical purposes, has been assumed here. 

To conclude, a mean-field solution for a kinetic exchange model of opinion formation and its phase transition in terms of forming global consensus 
is presented here. A genrelization to also include the three-agent interactions is proposed.
Surprisingly, the nature of the transition depends on the relative probabilities of the two-agent and three-agent interactions.     

\appendix
\section{Derivation of the polarization condition $f_1f_{-1}=0$}
\noindent Here we show that for the models with two-agent interactions studied here, in the steady state of the ordered phase agents with opinion of opposite
signs do not coexist in both `space' (all agents) and time. In other words, we show here that $f_1^xf_{-1}^y=0$ for any $x,y \ne 0$. 

For the model defined by Eq.~(\ref{lccc}), the time evolution equations governing $f_1$ and $f_{-1}$ are
\begin{eqnarray}
\frac{df_{-1}}{dt} = \frac{f_0f_{-1}p}{2}-f_{-1}^2(1-p)&&-f_{-1}f_1(1-\frac{p}{2})\nonumber \\
&& -f_{-1}f_0(1-p), \nonumber \\
\end{eqnarray}
\begin{eqnarray}
\frac{df_{1}}{dt} = \frac{f_0f_1p}{2}-f_1^2(1-p)&&-f_1f_{-1}(1-\frac{p}{2})\nonumber \\
&& -f_1f_0(1-p).
\end{eqnarray}
In the steady state we should have
\begin{eqnarray}
\frac{f_0f_{-1}p}{2}-f_{-1}^2(1-p)&&-f_{-1}f_1(1-\frac{p}{2})\nonumber \\ &=& f_{-1}f_0(1-p), 
\end{eqnarray}
\begin{eqnarray}
\frac{f_0f_1p}{2}-f_1^2(1-p)-&&f_1f_{-1}(1-\frac{p}{2}) \nonumber \\ &=& f_1f_0(1-p).
\end{eqnarray}
Adding the above two equations and imposing the condition for order state ($f_1\ne f_{-1}$) we get
\begin{eqnarray}
&& (1-p)(1-f_0)^2  -2f_1f_{-1}(1-p)+2f_1f_{-1}(1-\frac{p}{2})\nonumber \\ && +f_0(1-p)(1-f_0) 
=\frac{f_0p}{2}(1-f_0).
\end{eqnarray}
Now, it is already known without assuming $f_1f_{-1}=0$ [because the terms containing $f_1f_{-1}$ were canceled in Eq.~(\ref{opdt})] 
that in the ordered state $f_0=2(1-p)/p$. Using this in the above equation and simplifying, we get
\begin{equation}
f_1f_{-1}=0.
\end{equation}

In the generalised version (Eq.~(\ref{ps})) the time evolution equations for $f_1$ and $f_{-1}$ will take the form
\begin{eqnarray}
\frac{df_1}{dt} &=& \frac{f_0f_{1}q}{2}-f_1^2(1-p)(1-\frac{q}{2})-f_1f_{-1}\{(1-\frac{pq}{2})\nonumber \\ &&+(1-p)(1-\frac{q}{2})\}-f_1f_0(1-p),  \\
\frac{df_{-1}}{dt} &=& \frac{f_0f_{-1}q}{2}-f_{-1}^2(1-p)(1-\frac{q}{2})-f_1f_{-1}\{(1-\frac{pq}{2})\nonumber \\ && +(1-p)(1-\frac{q}{2})\}-f_{-1}f_0(1-p).
\end{eqnarray}
For steady state, the left hand sides will be zero. Then, as before, adding the resulting two equations and simplifying we get
\begin{eqnarray}
&& (1-p)(1-\frac{q}{2})\{(1-f_0)^2-2f_1f_{-1}\}+2f_1f_{-1}\{(1-\frac{pq}{2})\nonumber \\ &&+(1-p)(1-\frac{q}{2})\}+f_0(1-f_0)(1-p)=\frac{f_0q}{2}(1-f_0).
\end{eqnarray}
Again, it was derived (without assuming polarization) that in the ordered state $f_0=(p-1)(q-2)/pq$. Using this in the above equation and simplifying
one would get $f_1f_{-1}=0$.

So we see that in the present model (and in its generalised version), in the ordered state, $f_1^xf_{-1}^y=0$ for all $x,y \ne 0$ as was used in the main text.
Note that once one of the two extremes ($-1$ or $+1$) are completely eliminated, it can not come back. So the non-coexistence is valid not only for `space'
but also for all times (after the steady state is reached). 
Note, however, that this condition ($f_1f_{-1}=0$) was assumed in Sec. IV and V and were numerically verified. 
\begin{acknowledgements}
\noindent The author acknowledges stimulating discussions and suggestions of B. K. Chakrabarti , A. Chatterjee and S. Dasgupta. 
The author also thanks D. Bagchi and A. Ghosh for critical reading of the manuscript.
\end{acknowledgements} 
 %%%%%%%%%%%%%%%%%%%%%%%%%%%%%%%%%%%%%%%%%%%%%%%%%%%%%%%%%%%%%%%%%%%%

%%%%%%%%%%%%%%%%%%%%%%%%%%%%%%%%%%%%%%%%%%%%%%%%%%%%%%%%%%%%%%%%%%%
\end{document}